\documentclass[traditabstract]{aa}
\usepackage{natbib,graphicx,hyperref} %,amsmath,amssymb,graphicx,hyperref}
\usepackage[varg]{txfonts}
\begin{document}

\title{Loss of toroidal magnetic flux by emergence of bipolar magnetic regions}
\author{R.~H. Cameron \and M. Sch{\"u}ssler}

\institute{Max-Planck-Institut f\"ur Sonnensystemforschung, Justus-von-Liebig-Weg 3, 
  37077 G{\"o}ttingen, Germany}
%\email{cameron@mps.mpg.de}
\date{Received ; accepted} 
\abstract{ 
  The polarity of the toroidal magnetic field in the solar convection zone periodically
  reverses in the course of the 11/22-year solar cycle. Among the
  various processes that contribute to the removal of `old-polarity'
  toroidal magnetic flux is the emergence of flux loops forming
  bipolar regions at the solar surface. We quantify the loss of
  subsurface net toroidal flux by this process. To this end, we
  determine the contribution of an individual emerging bipolar loop
  and show that it is unaffected by surface flux transport after
  emergence. Together with the linearity of the diffusion process this
  means that the total flux loss can be obtained by adding the
  contributions of all emerging bipolar magnetic regions. The
  resulting total loss rate of net toroidal flux amounts to $1.3\times
  10^{15}\,$Mx s$^{-1}$ during activity maxima and $6.1\times
  10^{14}\,$Mx s$^{-1}$ during activity minima, to which ephemeral
  regions contribute about 90\% and 97\%, respectively. This rate is consistent
  with the observationally inferred loss rate of toroidal flux into
  interplanetary space and corresponds to a decay time of the
  subsurface toroidal flux of about 12~years, also consistent with a
  simple estimate based on turbulent diffusivity. Consequently,
  toroidal flux loss by flux emergence is a relevant contribution to
  the budget of net toroidal flux in the solar convection zone.
  That the toroidal flux loss rate due to flux emergence is consistent with
      what is expected from turbulent diffusion, and that the corresponding
      decay time is similar to the length of the solar cycle are important constraints
for understanding the solar cycle and the Sun's internal dynamics.}
 \authorrunning{Cameron et al.}  \titlerunning{Loss of
  toroidal magnetic flux} \maketitle

\section{Introduction}
\label{sec_intro}

The solar dynamo consists of poloidal magnetic field being wound up to generate
toroidal magnetic field, while a process involving the Coriolis force
creates poloidal field from the toroidal field \citep[see reviews by,
  e.g.][]{Ossendrijver:2003, Charbonneau:2010, Charbonneau:2014,
  Cameron:etal:2017, Brun:Browning:2017}. The oscillatory nature of
the solar cycle together with Hale's polarity rules
\citep{Hathaway:2015} imply a polarity reversal of the toroidal flux
system within the convection zone during each 11-year
cycle. Therefore, the question arises as to how the `old-polarity'
toroidal flux is disposed of before it is replaced by the
`new-polarity' flux. Principally, four different mechanism could
contribute:
\begin{enumerate}
\item{`Unwinding' by the action of differential rotation on the
  new (reversed) poloidal field,}
\item{cancellation of opposite-polarity magnetic flux at the
  equatorial plane due to latitudinal transport of toroidal flux by
  meridional flow, turbulent diffusion/pumping, or dynamo wave
  propagation \citep[e.g.,][]{Cameron:Schuessler:2016},}
\item{O-type neutral point dissipation along the dipole axis,}
\item{loss through the surface due to flux emergence.}
\end{enumerate}
    The first two possibilities are discussed in \cite{Wang:Sheeley:1991},
        as is the fourth possibility which they discount for the same reasons
        as put forward by  
%How flux emergence could provide loss of net toroidal flux has been
%discussed by
\citet{Parker:1984} and by
\citet{Vainshtein:Rosner:1991}. These authors pointed out that
(nearly) perfect flux freezing implies the necessity of detaching the
magnetic field lines from their mass load in order to be able to
escape from the solar interior. Flux emergence in the form of loops
could provide a path to such escape through a well-organized sequence
of reconnection events between adjacent ('sea-serpent') loops. Such a
situation, however, is considered to be rather artificial and 
in fact is not supported by observations.
The last process has also been considered as a nonlinearity limiting
the amplitude of the dynamo process \citep[e.g.,][]{Leighton:1969,
Schmitt:Schuessler:1989}.

In this paper, we consider the problem of toroidal flux loss by flux
emergence from a somewhat different perspective. We consider the net
toroidal flux integrated over a hemispheric meridional section, $\int
\langle B_\phi(r,\theta)\rangle\,dS$, where $\langle B_\phi\rangle$ is
the azimuthally averaged magnetic field
\citep{Cameron:Schuessler:2015}. For a reduction of $\langle
B_\phi\rangle$ and thus of the net toroidal flux it is not required
that toroidal field lines completely detach from the solar interior:
each single emergence of a loop reduces $\langle B_\phi\rangle$
proportional to the width and the flux of the loop. We show that the
contribution of each emerged loop to the reduction of the net toroidal
flux remains constant during the subsequent evolution of the emerged
flux. Therefore, the total amount of flux loss can be estimated by
simply adding up the contributions of all flux emergence
events. Eventually, the corresponding amount of toroidal flux is
carried away from the Sun by the solar wind and coronal mass ejections
\citep{Bieber:Rust:1995}.

The paper is organized as follows. In Sec.~\ref{sec_contr} we consider
the evolution of the net toroidal flux using the procedure developed
by \citet{Cameron:Schuessler:2015}. In Sec.~\ref{sec_loss} we discuss
the effect of loop emergence and the subsequent surface evolution of
flux on the net hemispheric toroidal flux. Quantitative estimates for
the resulting loss of net toroidal flux on the basis of observed
emergence rates are determined and compared with a simple estimate
based on turbulent diffusion. Sec.~\ref{sec_concl} contains our
conclusions.

\section{Evolution of the net toroidal flux}
\label{sec_contr}

\begin{figure}
\begin{center}
\includegraphics[scale=0.35]{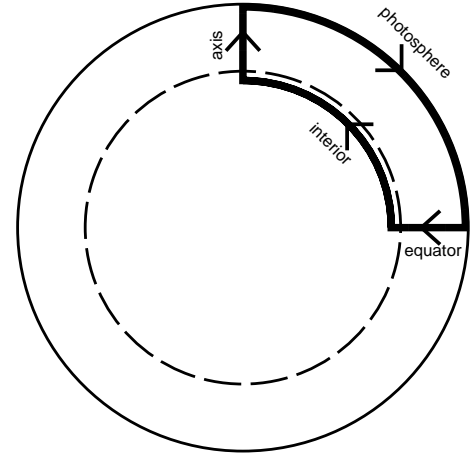}
\caption{Contour relevant for determining the evolution of the net
  toroidal flux in the northern hemisphere. The dashed line represents the
base of the convection zone.}
\label{fig:contour}
\end{center}
\end{figure}

Hale's polarity rules and the observation that the azimuthal field at
the solar surface shows a latitude-independent east-west orientation
in each hemisphere during the periods of maximum activity
\citep{2018A&A...609A..56C} suggest that the relevant quantity for the
large-scale solar dynamo is the net hemispheric toroidal flux in the
convection zone. \citet{Cameron:Schuessler:2015} have shown that
the evolution equation for the net toroidal flux in the (say) northern
hemisphere, $\Phi(t)$, is obtained in terms of a contour
integral by integrating the hydromagnetic induction equation
(neglecting the molecular diffusivity) over a hemispheric meridional
section of the convection zone and applying Stokes' theorem
(see Fig.~\ref{fig:contour}), viz.
\begin{equation}
  \frac{{\mathrm{d}} \Phi}{{\mathrm{d}}  t}
  =\oint ({\bf{U}}\times {\bf{B}})  \cdot \mathrm{d}{\bf l} =
   \oint (\langle{\bf{U}}\rangle\times \langle{\bf B}\rangle +
             \langle {\bf{U'}}\times {\bf B'}\rangle) \cdot \mathrm{d}{\bf l}\,.
\label{eq:stokes}
\end{equation}
Here ${\bf U}$ is the velocity field and ${\bf B}$ is the magnetic
field.  Quantities in angular brackets, $\langle\dots\rangle$, are
azimuthal averages and primed quantities represent fluctuations with
respect to the average. This equation describes both the generation of
net flux by differential rotation and the loss of flux.  In
particular, flux loss by transport through the surface is included in
the surface part of the contour integral,
\begin{eqnarray}
  \left(\frac{\mathrm{d} \Phi}{\mathrm{d} t}\right)_{\rm surf} 
   =\int^{\pi/2}_0 && \left( \langle U_\phi \rangle \langle B_r \rangle
   - \langle U_r \rangle \langle B_\phi \rangle \right. \nonumber \\
   &&+ \left. \langle U'_\phi B'_r \rangle -
   \langle U_r' B'_\phi \rangle  \right)\vert_{R_\odot} R_{\odot}  
   \mathrm{d}\theta\,, 
\label{eq:surf}
\end{eqnarray}
in spherical polar coordinates.  The first term of the integrand,
$\langle U_\phi \rangle \langle B_r \rangle$, represents the effect of
winding/unwinding by latitudinal differential rotation. The second
term, $\langle U_r \rangle \langle B_\phi \rangle$, vanishes since
there is no mean radial flow at the solar surface. The third term,
$\langle U'_\phi B'_r \rangle$, is negligible apart during flux
emergence events. This is because the evolution of the surface flux
after emergence is well represented by passive transport independent
of magnetic polarity, as demonstrated by the success of surface flux
transport simulations in reproducing the observations
\citep{Wang:etal:1989,Whitbread:etal:2017,Jiang:etal:2014,Jiang:etal:2015}.
The fourth term, $\langle U_r' B'_\phi \rangle$, represents flux
emergence and submergence. The latter process takes place 
when magnetic features of opposite polarities meet and cancel after reconnection.

\section{Flux loss by flux emergence}
\label{sec_loss}

As we have seen above, flux emergence and submergence change
    the net toroidal magnetic flux in the convection zone. The
    evolution of bipolar regions after emergence is well described as
    passive flux transport by horizontal flows (differential rotation,
    meridional circulation, and convective flows described in terms of
    turbulent diffusion). Submergence is represented by diffusive flux
    cancellation at locations where opposite polarities meet and one
    might therefore expect that the amount of toroidal flux loss 
    changes in the course of the evolution of a bipolar
    region. However, we show in this section that this is not the case
    and, furthermore, that the total amount of flux loss can be
    quantitatively estimated by simply adding the contributions of all
    bipolar magnetic regions.

\begin{figure}
\begin{center}
\includegraphics[scale=0.6]{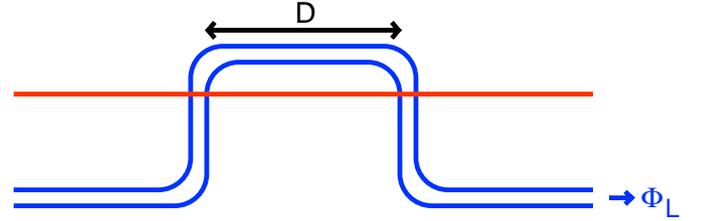}
\caption{Idealized sketch of an emerged loop of toroidal magnetic
  flux.  The flux tube (in blue) has a magnetic flux of $\Phi_L$; the
  longitudinal extent of the emerged part is $D$. The photosphere is
  indicated by the red line.}
\label{fig:loop}
\end{center}
\end{figure}

Consider a single emerged loop of toroidal magnetic flux as sketched
in Fig.~\ref{fig:loop}. The resulting decrease of the azimuthal
average, $\langle B_\phi \rangle$, corresponds to a reduction,
$\Delta\Phi_{\rm tor}$, of the subsurface net toroidal flux, given by
\begin{equation}
\Delta\Phi_{\rm tor} = \frac{D\, \Phi_{\rm L}}{2\pi R_\odot\cos\lambda}\,,
\label{eq:loop}
\end{equation}
where $\Phi_{\rm L}$ is the amount of flux contained in the loop, $D$
is the longitudinal extension of the loop after emergence, and
$\lambda$ its position in latitude. Equation~(3)  follows from the fact that the 
    azimuthally averaged change in the subsurface toroidal flux due to the emergence is
    equal to the flux of the loop multiplied by the fraction of the longitudinal separation of the
    two polarities at the surface to the circumference of the Sun at the latitude of emergence.
    More simply, the change in the longitudinally averaged subsurface toroidal flux due to an emergence
    is the flux of the emerging flux tube multiplied by the fraction of the tube in longitude which has
    moved across the photosphere.  
Various processes can, in
principle, affect the subsequent evolution of the emerged flux
contained in the corresponding bipolar magnetic region (BMR):
\begin{enumerate}
\item{Transport by horizontal convective flows, which can be described
  as turbulent diffusion \citep[random walk, see][]{Leighton:1964},}
\item{latitudinal differential rotation acting on tilted BMRs,}
\item{meridional flow, and}
\item{longitudinal drift of the two polarities in opposite directions caused
by magnetic tension in the subsurface part of the loop \citep{van_Ballegooijen:1982}.}
\end{enumerate}
Surface flux transport simulations have repeatedly demonstrated that
the surface magnetic flux is passively transported by the surface
flows, so that the fourth (dynamic) process seems irrelevant for the
evolution of the net toroidal magnetic flux.  While Eq.~\ref{eq:surf}
implies that meridional flow does not affect the net toroidal magnetic
flux, latitudinal differential rotation leads to the `unwinding' and
eventual reversal of the net toroidal flux in the course of the dynamo
process.

The question remains how far the first process, horizontal turbulent
diffusion, which causes cancellation, dispersal, and reconnection of
the emerged surface flux, leads to a temporal change of the amount of
flux loss given by Eq.~(\ref{eq:loop}). The relevant properties of the
process in this regard are that (1) diffusion is symmetric
(independent of polarity), i.e., it affects both polarities of the
loop flux in the same way, and (2) diffusion is a linear process, so
that the effects of many BMRs can be simply determined by adding
together the contributions of the individual BMRs, thus automatically
taking account of the permanent reorganisation of the surface field by
reconnection and cancellation of magnetic flux. It therefore suffices
to solely consider the evolution of one loop. 

In the course of the diffusive evolution, both opposite-polarity
patches of the vertical loop flux spread in all horizontal directions.
While Eq.~(\ref{eq:surf}) shows that expansion in latitude does not
affect the subsurface net toroidal flux, spreading in the longitudinal
direction potentially could. Part of the emerged flux cancels at the
neutral line between the polarities and thus 'heals' the
subsurface toroidal flux. Another part of the flux expands longitudinally
away from the neutral line and
eventually diffuses all around the Sun, thus finally removing the
corresponding amount of toroidal flux.  While the cancellation at the
neutral line reduces the amount of flux loss, the expanding part
increases the flux loss by effectively enlarging the polarity
separation, $D$. Owing to flux conservation and the symmetry of the
diffusion process, it turns out that both contributions exactly balance
each other, so that the loss of net toroidal flux, $\Delta\Phi_{\rm
  tor}$, remains time-independent at its initial value given by
Eq.~(\ref{eq:loop}).  This can be seen formally by the following
illustrative calculation.

Assume, for simplicity, one-dimensional cartesian geometry with a
purely vertical field, $B(x,t)$, that depends on the horizontal
coordinate, $x$ (representing the longitudinal direction), 
and time, $t$, in an infinite domain.  Consider the
evolution by diffusion of a bipolar region of vertical flux that is
centred at $x=0$ with the two polarities centred at $x=\pm x_0$. The
evolution of both polarities can be described by the analytical
solution for the diffusive spread of an initial delta function in
terms of Gaussian profiles, viz.
\begin{equation}
B(x,t) = B_0 \sqrt{\frac{a}{\pi}}
         \left[ e^{-a(x-x_0)^2}
          -   e^{-a(x+x_0)^2} \right] \,,
\label{eq:gauss}
\end{equation} 
with $a=(4\eta t)^{-1}$
and diffusivity $\eta$. The centre of gravity of the
field distribution for $x\geq 0$ is given by
\begin{equation}
{{\overline{x}}}_+ = \frac{\int_0^\infty Bx\,{\mathrm d}x}{\int_0^\infty B\, {\mathrm d}x} \,.
\label{eq:cog}
\end{equation} 
The centre of gravity, ${{\overline{x}}}_-$ for $x\leq 0$ is defined
analogously. The symmetry of the configuration entails
${{\overline{x}}}_- = -{{\overline{x}}}_+$.  The relevant quantity for
the reduction of the subsurface horizontal flux, corresponding to
$D\,\Phi_{\rm L}$ in Eq.~(\ref{eq:loop}), is given by
\begin{equation}
R(t) = {\overline{x}}_+ \int_0^\infty B\, {\mathrm d}x +
       {\overline{x}}_- \int_{-\infty}^0 B\, {\mathrm d}x 
     = 2 \int_0^\infty Bx\, {\mathrm d}x \,,
\label{eq:red}
\end{equation}
again owing to symmetry. 
Using Eq.~(\ref{eq:gauss}) we obtain
\begin{equation}
\int_0^\infty Bx\, {\mathrm d}x = B_0\sqrt{\frac{a}{\pi}}\, (I_+ - I_-)
\label{eq:int12a}
\end{equation} 
with
\begin{eqnarray}
I_+ &=& \int_0^\infty x e^{-a(x-x_0)^2} {\mathrm d}x \nonumber \\
I_- &=& \int_0^\infty x e^{-a(x+x_0)^2} {\mathrm d}x \,.
\label{eq:int12b}
\end{eqnarray} 
After some elementary algebra we obtain
\begin{equation}
(I_+ - I_-) = 2 x_0 \int_{x_0}^\infty e^{-a(x-x_0)^2} {\mathrm d}x 
            = x_0 \sqrt{\frac{\pi}{a}} \,,
\label{eq:int12c}
\end{equation}
so that with Eq.~(\ref{eq:red}) we have
\begin{equation}
R(t) = 2B_0\sqrt{\frac{a}{\pi}}\cdot x_0 \sqrt{\frac{\pi}{a}} = 2B_0 x_0\,,
\label{eq:red2}
\end{equation}
which is independent of time. That means that the diffusive evolution
of a bipolar magnetic region does not change the reduction of the
net toroidal flux due to its emergence, which is given by
Eq.~(\ref{eq:loop}). The increase of flux loss by the outward spreading
of the magnetic flux at the surface is exactly balanced by flux cancellation.
In fact, this result does not depend on the special assumption of 
Gaussian profiles but is valid for any symmetric profile that is
uniformly stretched while keeping the integral constant.

Since the flux loss, $\Delta\Phi_{\rm tor}$,
associated with an individual bipolar region is time-independent and
diffusion is a linear process, we can estimate the
mean rate of flux loss during a time interval $\Delta t$ by simply
adding the individual contributions given by Eq.~(\ref{eq:loop}) of
the bipolar regions emerging within that time, viz.
\begin{equation}
  \frac{{\rm d}\Phi_{\rm tor}}{{\rm d}t} = 
  \frac{\gamma \sum_i \left({D\,\Phi_{\rm L}}{(\cos\lambda)^{-1}}\right)_i}
       {2\pi R_\odot \Delta t}\,.
\label{eq:dphidt}
\end{equation}
The factor $\gamma$ is the fraction of the emerged flux that
    is not balanced by emergences with the opposite polarity
    orientation, i.e.,  $\gamma=(\Phi_{\rm Hale}-\Phi_{\rm
      non-Hale})/(\Phi_{\rm Hale}+\Phi_{\rm non-Hale})$, where
    $\Phi_{\rm Hale}$ and $\Phi_{\rm non-Hale}$, respectively, are the
    amounts of flux that emerge obeying Hale's law and not obeying
    it.
%This factor accounts for the fraction of bipolar regions not
%following Hale's polarity rules.

We first consider the contribution due to ephemeral regions, small
bipolar regions carrying a magnetic flux of the order of $10^{20}\,$Mx
that emerge ubiquitously at the solar surface.
\citet{2001ApJ...555..448H} determined a value of $5 \times
10^{23}$~Mx per day for the emergence rate of unsigned flux in
ephemeral regions over the entire solar surface.  About 60\% of these
were found to obey Hale's polarity laws (i.e., a surplus of 20\%), so
that $\gamma=0.2$ in this case. For a rough estimate of the
corresponding loss of toroidal flux we assume that polarity
separation, $D$, loop flux, $\Phi_{\rm L}$, and emergence latitude,
are all uncorrelated.  Since the contribution of each emerging loop to
the total unsigned surface flux equals $2\Phi_{\rm L}$, we have
$\sum_i \Phi_{\rm L}=2.5 \times 10^{23}$~Mx per day.  The average
polarity separation for ephemeral regions is about $9\,$Mm
\citep{2001ApJ...555..448H}. Since $D$ is the longitudinal separation,
we have $D_i=9 \cos(\alpha_i)$~Mm where $\alpha_i$ is the tilt angle
of the axis of the ephemeral region with respect to the east-west
direction. For a given longitudinal polarity orientation (Hale or
anti-Hale), these angles are likely to be uniformly distributed
between $\pm 90^{\circ}$, so that on average we expect $\langle D_i
\rangle \approx 9\times 0.64=5.76$~Mm where $0.64$ is the average
value of $\cos(\alpha)$ between $-90^{\circ}$ and $90^{\circ}$.  We
assume the emergences occur uniformly over the surface, so that the
weighted average of $\cos(\lambda)^{-1}$ over the emergences is
\begin{eqnarray}
  \langle (\cos\lambda)^{-1} \rangle_i &=& \frac{\int^{90^{\circ}}_0 
  \cos(\lambda)^{-1} \cos\lambda d\lambda}
  {\int^{90^{\circ}}_0 \cos\lambda d\lambda}
  \nonumber \\
&=&\pi/2\,,
\end{eqnarray}
where the weighting factor $\cos\lambda$ accounts for the fact that
the length of the circumference at constant latitude is proportional to
$\cos\lambda$.  We thus obtain for the loss rate of toroidal flux
per hemisphere due to the emergence of ephemeral regions a value of
\begin{equation}
  \frac{{\rm d}\Phi_{\rm tor, hem}^{\rm ER}}{{\rm d}t} \approx
  5.9 \times10^{14}\mbox{ Mx s}^{-1}\,.
\label{eq:dphidthER}
\end{equation}
The results of \citet{2001ApJ...555..448H} are based on data from
October 1997, i.e., under solar minimum conditions. Since the
emergence rate of ephemeral regions varies roughly by a factor of 2-3
during the solar cycle \citep{Harvey:etal:1975, Martin:Harvey:1979,2003ApJ...584.1107H},
we expect the loss rate during solar maxima to be correspondingly
higher, so that
\begin{equation}
  \frac{{\rm d}\Phi_{\rm tor, hem}^{\rm ER}}{{\rm d}t} \vert_{\rm maximum} \approx
  2\times 5.9 \times10^{14}\mbox{ Mx s}^{-1}\,.
\label{eq:dphidthER_max}
\end{equation}

We note that a few years before activity minima ephemeral regions from
the current and the next cycle are both present on the surface, and
the change in the net hemispheric subsurface toroidal flux will
reflect the difference between the contributions of the old and new
cycle ephemeral regions.

For active regions exceeding 3.5 square degrees in size,
\citet{1994SoPh..150....1S} report emergence rates over the
entire solar surface of $7.4 \times 10^{20}$~Mx per day during
activity minimum and $6.2 \times 10^{21}$~Mx per day during
maximum. For simplicity, we assume an average polarity separation of
$40\,$Mm, east-west alignment, and emergence close to the equator
$\cos\lambda=1$.  We then obtain
\begin{equation}
  \frac{{\rm d}\Phi_{\rm tor, hem}^{\rm AR}}{{\rm d}t} \approx
  2.0\times10^{13}\mbox{ Mx s}^{-1}
\label{eq:dphidthARmin}
\end{equation}
during activity minimum and
\begin{equation}
  \frac{{\rm d}\Phi_{\rm tor, hem}^{\rm AR}}{{\rm d}t} \approx
  1.6\times10^{14}\mbox{ Mx s}^{-1}
\label{eq:dphidthARmax}
\end{equation}
during maximum. Assuming a factor of 2 variation of the emergence rate
of ephemeral regions between minimum and maximum, these regions
therefore contribute about 90\% of the total loss rate of toroidal flux
during solar maxima and about 97\% during minima. 
The total flux loss rate per  hemisphere during minimum is then
\begin{equation}   
(\frac{{\rm d}\Phi_{\rm tor, hem}^{\rm ER}}{{\rm d}t}+
  \frac{{\rm d}\Phi_{\rm tor, hem}^{\rm AR}}{{\rm d}t})\vert_{\mathrm{minimum}}
                   =6.1\times10^{14} \mbox{~Mx~s}^{-1},
\end{equation}  
and  $1.3\times10^{15}$~Mx~s$^{-1}$ during maximum. 
With the total loss rate of $1.3\times10^{15}$ Mx s$^{-1}$ around
maxima and a total amount of subsurface toroidal flux of $5\times
10^{23}$~Mx per hemisphere \citep{Cameron:Schuessler:2015}, we obtain
a characteristic decay time of 12.2 years. Consequently, flux loss
through the photosphere associated with flux emergence is an important
factor for the evolution of the subsurface toroidal flux on
solar-cycle timescales. Roughly approximating the cycle-averaged loss
rate by the mean of its maximum and minimum values, i.e.,
$9.6\times10^{14}$ Mx s$^{-1}$, we obtain a total loss of toroidal
flux by flux emergence over 11~years of $3.3\times10^{23}\,$Mx.
\citet{Bieber:Rust:1995} estimated the total loss of toroidal flux
into interplanetary space as $10^{24}\,$Mx per 11-year cycle, i.e.,
$5\times10^{23}$~Mx per hemisphere and cycle. This is roughly
consistent with our result of $3.3\times10^{23}\,$Mx per hemisphere
and cycle, given the considerable uncertainties and simplifications
entering both estimates.

The rate at which BMRs appear on the solar surface, as a function of the
amount of flux which emerges, is described by a single power law which extends
over 5 orders of magnitude, from small ephemeral regions to large active regions
\citep[see, for example][]{2003ApJ...584.1107H, 2011SoPh..269...13T}. Unlike active regions
which emerge only at latitudes less than about 40$^{\circ}$, 
ephemeral regions emerge all over the solar surface. However 
ephemeral regions emerging in the butterfly wings have a tendency to obey Hale's law,
with the same east-west orientation as the active regions of the
same cycle \citep{1979SoPh...64...93M}. The tendency of ephemeral region
to emerge obeying Hale's law extends the butterfly wings to
earlier times and higher latitudes \citep{1979SoPh...64...93M, 1988Natur.333..748W}.
    
Which sizes range of BMRs are most important for the
loss of toroidal field through the surface is mainly decided by the
competition between the number of emergences and their tendency to obey Hale's law.
The ephemeral regions dominate the flux loss at all phases of the
solar cycle because ephemeral region emergence is much more common than
active region emergence. The larger ephemeral regions in the range of $10^{18}$~Mx
and above are presumably more important than the smaller emergences because
the tendency to obey Hale's law decreases rapidly with decreasing flux of the BMR 
\citep{2003ApJ...584.1107H}.

We can also compare our result with simple estimates in terms of
turbulent diffusion.  Instead of regarding individual emergence
events, this approach considers the transport of toroidal magnetic
field by turbulent motions throughout the convection zone and across
the photosphere. Ignoring turbulent pumping, one can parameterize this
by an effective turbulent diffusivity, $\eta_{\mathrm{t}}$, the value
of which can be estimated using mixing length theory
\citep[e.g.][]{2011ApJ...727L..23M}, from numerical simulations
\citep[e.g.][]{2018A&A...609A..51W}, or inferred from observations
\citep[e.g.][]{Cameron:Schuessler:2016}. Near-surface values of
$\eta_{\mathrm{t}}$ are typically around $10^{12}$cm$^2\,$s$^{-1}$.
Using the depth of the convections zone, $L=200$~Mm, as a typical
length scale, this leads to a diffusive decay time of
$\tau=L^2/\eta_{\mathrm{t}} \simeq 12.7\,$years, which is consistent
with the above value of 12.2 years from flux emergence.

\section{Conclusions}
\label{sec_concl}

Our results show that the loss of net toroidal flux from the solar
interior due to flux emergence can be faithfully estimated by adding
the time-independent contributions of the individual bipolar regions
to the reduction of the longitudinally averaged azimuthal field.
Using the observed emergence rates of ephemeral and active regions
leads to a characteristic decay time of the toroidal flux of about 12
years, in which ephemeral regions contribute most of the effect.
The decay rate of toroidal flux by flux emergence is also consistent
with simple estimates based on turbulent diffusion. Consequently,
flux emergence represents a relevant loss mechanism for the interior
toroidal flux.
The decay of toroidal flux is further enhanced by
cancellation across the equator, dissipation along the dipole axis,
and `unwinding' by differential rotation. However, these processes presumably
are not dominant because the toroidal flux loss through the photosphere
already accounts for most of what needs to be removed.
The 12-year timescale for toroidal flux loss due to flux emergence
is close to the 11-year solar cycle period. This means that the flux loss
is very important to the subsurface flux evolution, and the 12-year timescale
is an important constraint for models of the solar dynamo.

\begin{acknowledgements}
 RHC  acknowledges  partial  support  from  ERC  Synergy  grant  WHOLE  SUN  810218.  
\end{acknowledgements}
\bibliographystyle{aa}
\bibliography{TOR}

\end{document}